\documentclass[prd,showpacs]{revtex4}
\usepackage{graphicx}
\usepackage{amsmath}
\usepackage{amssymb}
\begin{document}

\renewcommand{\theequation}{\thesection.\arabic{equation}}

\newcommand{\re}{\mathop{\mathrm{Re}}}

\newcommand{\be}{\begin{equation}}
\newcommand{\ee}{\end{equation}}
\newcommand{\bea}{\begin{eqnarray}}
\newcommand{\eea}{\end{eqnarray}}

\title{Randall-Sundrum limit of $f(R)$ brane-world models}

\author{Adam Balcerzak}
\email{abalcerz@wmf.univ.szczecin.pl}
\author{Mariusz P. D\c{a}browski}
\email{mpdabfz@wmf.univ.szczecin.pl}
\affiliation{\it Institute of Physics, University of Szczecin, Wielkopolska 15, 70-451 Szczecin, Poland}
\affiliation{\it Copernicus Center for Interdisciplinary Studies,
S{\l }awkowska 17, 31-016 Krak\'ow, Poland}

\date{\today}

\input epsf

\begin{abstract}

By setting some special boundary conditions in the variational principle we obtain junction conditions for the five-dimensional 
$f(R)$ gravity which in the Einstein limit $f(R)\rightarrow R$ transform into the standard Randall-Sundrum junction conditions. 
We apply these junction conditions to a particular model of a Friedmann universe on the brane and show explicitly that the limit 
gives the standard Randall-Sundrum model Friedmann equation.

\end{abstract}


\maketitle

\section{Introduction}

Brane universes are now one of the most interesting option for unifying gauge interactions with gravity at the TeV scale \cite{brane1}. 
They were nicely introduced by Randall and Sundrum \cite{RS} and further developed cosmologically \cite{brane2}. However, it is not trivial 
to combine brane theories with higher-order gravity theories such as $f(R)$ or $f(R, R_{ab}R^{ab}, R_{abcd}R^{abcd})$ 
gravities \cite{f(R),faraoni10,NOreview} which are theories with lagrangians dependent on the curvature invariants. The obstacles are the 
ambiguities of the quadratic delta function contributions to the field equations. The only cases which naturally avoid these ambiguities are 
curvature invariants combinations which form Lovelock densities \cite{lovelock,GBbrane}. However, due to some new approaches 
(improvement of the continuity properties of the metric on the brane or reduction to a second-order theory) the obstacles were challenged 
successfully in Ref. \cite{paper1+2} and the Israel junction conditions \cite{israel66} were obtained following earlier discussion of Refs. 
\cite{borzeszkowski,deeg,hinterb,NOlinear}. The method of reduction $f(R)$ theory to a second-order theory is possible by the introduction 
of an extra scalar field -- the scalaron. In this approach $f(R)$ theory becomes the scalar-tensor Brans-Dicke gravity \cite{bd} with a 
Brans-Dicke parameter $\omega =0$, and the induced scalaron potential with the scalaron playing the role of the Brans-Dicke field. 
The junction conditions obtained in Ref. \cite{paper1+2} generalized both the junction conditions obtained in Refs. \cite{BDbrane,BDbranenoZ2} 
for the Brans-Dicke field without a scalar field potential, and also the conditions derived in Ref. \cite{deeg,deruelle07,afonso} for $f(R)$ 
brane gravity. In Refs. \cite{paper3,adamADP} previously derived junction conditions in \cite{paper1+2} were applied to cosmology.

It is important to say that the junction conditions for $f(R)$ gravity given in  \cite{paper1+2,paper3} were obtained by using the boundary 
conditions that assumed unrestricted variation of the metric and the scalaron on the brane. In fact, the freedom of the variation of the scalaron on 
the brane leads to the continuity of trace of the extrinsic curvature on it (see Eq. (2.7) in \cite{paper3}).  This property makes 
it impossible to obtain the standard Israel junction conditions in the Einstein limit $f(R)\rightarrow R$. However, the boundary 
conditions used in Refs. \cite{paper1+2,paper3} are not unique, and can be replaced by some different ones -- those which allow 
the Einstein limit and this is the task of the present paper.

In Section II we discuss the variational principle for $f(R)$ brane models reduced to second-order Brans-Dicke models and we derive 
the junction conditions which allow the Einstein limit. In Section III we apply these junction conditions to the Friedmann geometry on the brane. 
In Section IV we solve for the bulk anti-de-Sitter geometry, and in Section V we explicitly show how the proposed junction conditions work in 
the Einstein-Randall-Sundrum limit. In Section VI we give our conclusions.

\section{Boundary conditions and junction conditions with an RS limit}
\label{fRgrav}
\setcounter{equation}{0}

We start with the action of the $f(R)$ theory \cite{f(R)} in five dimensions which reads as
\bea
\label{ac}
S_{p} &=&  \frac{1}{2\kappa_5^2} \int_{M_p} d^5 x \sqrt{-g} f(R)+ S_{bulk,p}~,
\eea
where $R$ is the Ricci scalar, $\kappa_5^2$ is a five-dimensional Einstein constant, $S_{bulk,p}$ is the bulk matter action, 
and $M_p$ ($p=1,2$) is the spacetime volume. Since the action (\ref{ac}) gives fourth-order field equations, then it is advisable 
to use an equivalent action
\begin{eqnarray}
\label{equiv22}
\bar{S}_{p} &=& \int_{M_p}d^5 x\sqrt{-g}\{f'(Q)(R-Q) + f(Q)\}
\end{eqnarray}
where $Q$ is an extra field which plays the role of a Lagrange multiplier, and $f'(Q) \equiv
df(Q)/dQ$. Varying the action (\ref{equiv22}) with respect to $g_{ab}$ and $Q$ we obtain the equations of motion
\begin{eqnarray}
\label{fe}
\frac{1}{2} g^{ab} f(Q) -f'(Q) R^{ab} -g^{ab} f'(Q) + f'(Q)^{;ab}=0 ~~,\\
Q=R~~.
\end{eqnarray}
with the condition that $f''(Q) \neq 0$, and we interpret $f'(Q)$ as an extra scalar
field - the scalaron:
\be
\label{scalaron}
\phi = f'(Q) = f'(R)~.
\ee
Using the scalaron, the action (\ref{equiv22}) can be rewritten in the form of a Brans-Dicke action with a Brans-Dicke parameter $\omega=0$, i.e.,
\begin{eqnarray}
\label{equiv3}
\bar{S}_{p} &=& \int_{M_p} d^5 x \sqrt{-g}\{\phi R - V(\phi)\} + S_{bulk,p} .
\end{eqnarray}
where $V(\phi) = -\phi R(\phi) + f(R(\phi))$ \cite{paper1+2}.

In our previous papers \cite{paper1+2,paper3} we derived junction conditions for braneworld model action (\ref{ac}), but none of them 
possessed a standard Randall-Sundrum limit. Here we suggest an approach which allows to do so at the expense of choosing some special 
boundary conditions while varying the braneworld action with an appropriate Hawking-Lutrell boundary term which 
for the action (\ref{equiv3}) is \cite{lutrell}
\begin{equation}
\label{HL}
S_{HL_{p}}=-2(-1)^{p}\epsilon \int_{\partial M_p} \sqrt{-h} \phi K d^4x~,
\end{equation}
where $K$ is the trace of the extrinsic curvature tensor $K_{ab}$, $h$ is the determinant of the induced metric
$h_{ab} = g_{ab} - \epsilon n_an_b$, $n^a$ is a unit normal vector to a boundary $\partial M_p$, and
$\epsilon = 1$ $(\epsilon = -1)$ for a timelike (a spacelike) brane, respectively.
The total action of the theory is then
\begin{eqnarray}
\label{equiv4}
\bar{S}_{tot_{p}} &=& \bar{S}_{p}+S_{HL_{p}}.
\end{eqnarray}
The variation of the action (\ref{equiv4}) leads to
\begin{eqnarray}
\label{Vstotp}
\delta S_{tot_p} &=& \nonumber \\ \nonumber
&=& - \int_{\partial M_p}d^{4}x \sqrt{-h}(-1)^{p} \left\{ \epsilon
\left[(g^{ab}+ \epsilon n^{a}n^{b}) (\phi_{;e}n^e) \right. \right.
\\  \nonumber &+& \left. \left. 2 n^{b}h^{ea}\phi_{,e} + \phi K h^{ab} -  \phi K^{ab}
- 2n^{(a} \phi^{,b)} \right] \right. \delta g_{ab}\\
&-& 2(-1)^{p} \epsilon  \int_{\partial M_p}d^{4}x \sqrt{-h}
K \delta \phi~,
\end{eqnarray}
where the bulk parts have been omitted. Full variation over the bulk space,
separated by a brane, requires the variation of both of these parts
separately (i.e. first for $p =1$, and then for $p = 2$). This means that the full action
is
\begin{eqnarray}
\label{S12}
\bar{S}_{tot} =S_{tot_1}+S_{tot_2} + S_{brane}~,
\end{eqnarray}
where
\begin{equation}
\delta S_{brane} ={\kappa_{5}^2} \int_{\partial M_p} d^{4}x \sqrt{-h} S^{ab} \delta
g_{ab}~,
\end{equation}
and $S_{ab}$ is an energy-momentum tensor of the matter on the brane. 

Now, there is a crucial point which allows to obtain the Randall-Sundrum limit of the $f(R)$ brane junction conditions. 
Namely, we vary the total action (\ref{S12}) in such a way that we set boundary conditions as follows
\be
\label{bc}
\delta \phi \rightarrow 0 \hspace{0.5cm} {\rm for} \hspace{0.5cm} w\rightarrow 0~,
\ee
where $w$ is a coordinate normal to the brane. The physical meaning of (\ref{bc}) is that we impose the variation of the scalaron 
to vanish on the brane but not in the bulk. In fact, such a choice allows to kill the last term in the variation (\ref{Vstotp}) 
leaving the other terms untouched (since we did not assume that $\delta g_{ab} = 0$ on the brane) and also leaving the freedom of 
a choice for the trace of the extrinsic curvature which not necessarily has to be continuous on the brane. Such a choice of the boundary conditions 
allows to obtain the following junction conditions \cite{paper3}:
\begin{eqnarray}
\label{jc}
&-&(g^{ab}+ \epsilon n^{a}n^{b}) [\phi_{;c}n^{c}] -
 2 n^{(a}h^{eb)}[\phi_{,e}] \\ \nonumber &-& [\phi K] h^{ab} +  [\phi K^{ab}] + 2n^{(a} [\phi^{,b)}] =
 \epsilon{\kappa_{5}^2} S^{ab}~,
\end{eqnarray}
where $[A]\equiv A^+ - A^-$ for any quantity $A$ with its values $A^+$ and $A_-$ right and left of the brane, respectively.

Projecting  Eq. (\ref{jc}) onto the directions tangent to the brane by multiplying it by $h_{ac}h_{bd}$, we obtain:
\begin{eqnarray}
\label{torow}
-h_{ab}\left[\frac{\partial \phi}{\partial w} \right]-[\phi K]h_{ab} &+& [\phi K_{ab}]
= \epsilon {\kappa_{5}^2}S_{ab}.
\end{eqnarray}
Next, assuming that a brane is timelike ($\epsilon=1$), and that the scalaron is continuous on it, i.e. that
\be
[\phi]=0~,
\ee
and combining (\ref{torow}) with its contraction, we obtain the final junction conditions in a more convenient form 
\begin{equation}
\label{jcgut}
[K_{ab}]=\frac{\kappa_{5}^2 \left(S_{ab}-\frac{1}{3}h_{ab} S\right) - \frac{h_{ab}}{3}\left[\frac{\partial \phi}{\partial w}\right]}{\phi}.
\end{equation}
Finally, we can see that these $f(R)$ theory junction conditions (\ref{jcgut}) transform in the Einstein (or Randall-Sundrum) limit $f(R)\rightarrow R$, i.e., for (cf. (\ref{scalaron}))
\be
\label{philimit}
\phi \to 1, \hspace{0.3cm} \partial\phi/\partial w \to 0 
\ee
into the standard Israel junction conditions for a 5-dimensional spacetime \cite{brane2}:
\begin{equation}
\label{jcstan}
[K_{ab}]=\kappa_{5} \left(S_{ab}-\frac{1}{3}h_{ab} S\right).
\end{equation}
In Ref. \cite{paper3} we derived different type of junction conditions. Since the condition (\ref{bc}) was not imposed, then the scalar 
field part of the brane boundary term (\ref{Vstotp}) vanished provided that $[K]=0$, i.e., the trace of the extrinsic curvature was assumed 
to be continuous on the brane (Eqs. (2.7)-(2.10) of Ref. \cite{paper3}). After imposing a mirror symmetry the trace of the extrinsic 
curvature had additionally to vanish $K=0$ (Eqs. (2.11)-(2.13) of Ref. \cite{paper3}) which further under the requirement of vanishing the curvature 
scalar $R=0$ gave the special junction conditions obtained in Ref. \cite{deruelle07}, for example.

However, the junction conditions (\ref{jcgut}) are different and they allow the Randall-Sundrum limit which is quite beneficial for 
developing $f(R)$ cosmology on the brane.

\section{$f(R)$ Friedmann cosmology on the brane}
\label{fRcosm}
\setcounter{equation}{0}

In order to discuss Friedmann cosmology on the brane we start with a 5-dimensional spherically symmetric bulk metric given in the 
form \cite{mannheim} 
\begin{equation}
\label{sphersym}
ds^{2}=-h(r) dT^{2} + \frac{dr^{2}}{h(r)} +r^{2}d\Omega^{2}_{3},
\end{equation}
where
$$d\Omega^{2}_{3}=d\chi^{2} + sin^{2}\chi (d\theta^{2} + sin^{2}\theta d\phi^{2})$$
is the metric of a 3-dimensional unit sphere (which means that we assume the Friedmann curvature index $k=+1$ here). After making 
a coordinate transformation $T=T(w,\tau)$, $r=r(w,\tau)$, the metric reads as
\begin{eqnarray}
\label{mettran}
\nonumber
ds^{2}&=&\left\{-h(r)T'^{2} + \frac{r'^{2}}{h(r)}\right\}dw^{2} + \left\{{-h(r)\dot{T}^{2} + \frac{\dot{r}^2}{h(r)}}\right\}d\tau^{2} \\
&+&\left\{-2h(r) T'\dot{T} + \frac{2}{h(r)}r'\dot{r} \right\} dw d\tau + r^{2}d\Omega^{2}_{3}~~.
\end{eqnarray}
Now, we transform the metric (\ref{mettran}) into the form which defines the Gaussian normal coordinates \cite{mannheim}
\begin{eqnarray}
\label{gauss}
\nonumber
ds^{2}&=&dw^{2} - d\tau^{2} + r^{2}d\Omega^{2}_{3}~~,
\end{eqnarray}
by taking
\begin{eqnarray}
\label{gausscond}
-h(r)T'^{2} + \frac{r'^{2}}{h(r)}=1~~, \\
h^{2}(r) T'\dot{T} = r'\dot{r}~~,
\end{eqnarray}
(prime is the derivative with respect to $w$ and dot with respect to $\tau$) and further assuming that $\tau$ is a proper time on the brane i.e.
\begin{equation}
\label{propt}
{-h(r)\dot{T}^{2} + \frac{\dot{r}^2}{h(r)}}=-1~~.
\end{equation}
In order to calculate junction conditions (\ref{jcgut}) we need to calculate the components of the extrinsic curvature for 
the metric (\ref{gauss}), i.e., 
\begin{equation}
\label{ext}
K_{ab}=-\Gamma^{w}_{~~ab}=\frac{1}{2} \frac {\partial h_{ab}}{\partial w}~~,
\end{equation}
which after specifying the components of the induced metric $h_{ab}$
\begin{equation}
\label{ind}
h_{ab}=
\begin{vmatrix}
-1 & 0 & 0 & 0 \\
0 & r^{2} & 0 & 0  \\
0 &  0  & r^{2} sin ^{2} \chi & 0 \\
0 & 0 & 0 & r^{2} sin ^{2} \chi  sin ^{2} \theta
\end{vmatrix}
\end{equation}
allows the only non-vanishing term
\begin{equation}
\label{ext2}
K_{22}= r \frac {\partial r }{\partial w}=\pm r h(r) \dot {T}=\pm r \sqrt{\dot{r}^{2}+h(r)}~~.
\end{equation}

After assuming that there is a mirror symmetry ($A^+ = -A^- = A$, i.e. $[A] = 2A$ for any quantity $A$) and that the matter on the brane is 
in the form of a perfect fluid, the junction conditions (\ref{jcgut}) read as:
\begin{equation}
\label{be}
\pm 2r \sqrt{\dot{r}^{2}+h(r)}=\frac{\kappa_{5}^2 r^{2} \rho - \left[ \frac{\partial \phi}{\partial w} \right]r^{2}}{3\phi}~~.
\end{equation}

In order to proceed further and specify the function $h(r)$ we need to solve field equations in the bulk. 


\section{The solution of a vacuum 5-dimensional field equations for $f(R)$ theory}
\label{fR5dimsoln}
\setcounter{equation}{0}

The action (\ref{equiv22}) with $S_{bulk,p}=0$ for the spherically symmetric metric (\ref{sphersym}) can be expressed as
\begin{equation}
\label{equiv2}
\int r^3  sin ^{2} \chi cos \theta \left\{f(Q) - f'(Q)\left(Q + h''(r) + 6 \frac{h'(r)}{r} + 6 \frac{h(r)}{r^2} - \frac{6}{r^2}\right)\right\} d^{5}x~~,
\end{equation}
where for this metric \cite{sebastiani}
\begin{equation}
R= -h''(r) - 6 \frac{h'(r)}{r} - 6 \frac{h(r)}{r^2} + \frac{6}{r^2}~~.
\end{equation}
Varying the action (\ref{equiv2}) with respect to $h(r)$, one obtains (cf. Ref. \cite{sebastiani} - Eq. (15)) for the scalaron that
\begin{equation}
\label{constr}
f'(Q) = a r + b,
\end{equation}
where $a$ and $b$ are constants. Now, substituting (\ref{constr}) to the field equations (\ref{fe}) we obtain the differential equation 
for $h(r)$ as follows
\begin{equation}
\label{diff}
(ar+b) (r^{2} h''(r) + 4) + r (b +  2 a r) h'(r) -2(2b+3ar)h(r)=0.
\end{equation}
The most general solution of (\ref{diff}) is
\begin{eqnarray}
\label{sol}
&&h(r)=-\frac{a^{2}B}{2b^3}-\frac{B}{4br^2}+\frac{aB}{3b^{2}r}+\frac{a^{3}Br}{b^4}+Ar^{2}+\frac{a^{4} B r^{2} \ln(r)}{b^{5}}-\frac{a^{4}Br^{2}\ln(ar+b)}{b^5} \\ \nonumber
&-&\left(2br^{2}+ \frac{4ar^{3}}{3}\right)\left(-\frac{a^{2}}{2b^{3}}-\frac{1}{4br^{2}}+\frac{a}{3b^{2}r}+\frac{a^{3}r}{b^{4}}+\frac{a^{4}r^{2}\ln(r)}{b^5}
-\frac{a^{4}r^{2}\ln(ar+b)}{b^5}\right)\\ \nonumber
&+&\frac{b(3b^{4} - 2ab^{3}r+8a^{3}br^{3}+8a^{4}r^{4})+4a^{2}r^{2}(ar+b)^{2}(2ar-b)\ln(r)-4a^{2}r^{2}(ar+b)^{2}(2ar-b)\ln(ar+b)}{6b^{5}}~,
\end{eqnarray}
which in the case of a constant Ricci curvature $R=$ const. (i.e., $a=0$ and $f'(Q) =b$) gives the solution for a 5-dimensional anti-de-Sitter 
space in the form \cite{mannheim}
\begin{equation}
\label{ads}
h(r)=1-\frac{C}{4r^{2}} + A r^{2}~,
\end{equation}
with $A$, $B$ and $C=B/b$ being constants. 

\section{Randall-Sundrum limit of the vacuum $f(R)$ theory on the brane}
\label{fRRSlimit}
\setcounter{equation}{0}

In order to discuss a practical transition from the $f(R)$ junction conditions (\ref{jcgut}) to the Randall-Sundrum junction 
conditions (\ref{jcstan}) we first calculate a jump of the scalaron (\ref{constr}) as follows
\begin{eqnarray}
\label{tu}
 \nonumber
\left[  \frac{\partial \phi}{\partial w} \right]&=& \left[(a r(|w|,\tau)+b),_{w} \right]=a \frac{\partial r}{\partial |w|} \left[\frac{\partial |w|}{\partial w} \right] \\ &=& 2 a r'= 2 a h(r) \dot{T} =\pm 2a \sqrt{\dot{r}^{2}+h(r)}~~.
  \end{eqnarray}
Substituting (\ref{tu}) into (\ref{be}), we obtain the cosmological equation on the brane as
  \begin{equation}
  \label{eqbrane}
  \left(\frac{\dot{r}^{2} + h(r)}{r^2}\right)(4ar+3b)^{2}=\left(\frac{\kappa_{5}^2}{2}\right)^{2}\rho^{2}.
  \end{equation}
Finally, by taking the limit $a\rightarrow 0$ and $b\rightarrow 1$ 
(which is equivalent to the limit $\phi \to 1$, $\partial \phi/\partial w \to 0$ for (\ref{jcgut})), the equation (\ref{eqbrane}) takes the form
\begin{equation}
\label{RSSlimit}
\frac{\dot{r}^{2}}{r^{2}}= \frac{\kappa_{5}^{2}}{36}\rho^{2} - A + \frac{B}{4r^{4}} - \frac{1}{r^2}~,
\end{equation}
which further by assuming $A=-\Lambda_{5}/6$, and $B=4U$ becomes the cosmological equation of the well-known Randall-Sundrum-Friedmann 
braneworld model \cite{brane2}
\begin{equation}
\label{RSS}
\frac{\dot{r}^{2}}{r^{2}}= \frac{\kappa_{5}^{2}}{36}\rho^{2} +\frac{\Lambda_{5}}{6} - \frac{1}{r^2} + \frac{U}{r^{4}},
\end{equation}
where $\Lambda_{5}$ is the 5-dimensional bulk cosmological constant, and $U$ is the integration constant which refers to the dark radiation 
(note that the spatial curvature index is $k=+1$ here). In fact, the bulk cosmological constant is induced by geometry of $f(R)$ models despite 
that $S_{bulk,p}=0$ in the action (\ref{ac}). 

\section{Conclusions}

We have proposed a new type of brane models which are based on some special boundary conditions in the variational principle. Namely, while 
varying the Gibbons-Lutrell boundary term, we imposed the condition that the variation of the scalaron should vanish on the brane 
(but not in the bulk). This allowed to have less restrictive requirements related to the continuity of the trace of extrinsic curvature 
on the brane. Due to that we obtained junction conditions for the five-dimensional $f(R)$ gravity which in the Einstein limit 
$f(R)\rightarrow R$ transform into the standard Randall-Sundrum junction conditions. Further, we applied these junction conditions 
for a particular model of a Friedmann universe on the brane and show explicitly that the limit gives the standard Randall-Sundrum-Friedmann 
equation. The result is quite beneficial for developing $f(R)$ cosmology on the brane. 

\section{Acknowledgements}

We acknowledge the support of the National Science Center grant No N N202 3269 40 (years 2011-13). We are indebted to Bogus{\l }aw Broda, 
Salvatore Cappoziello, Krzysztof Meissner, Sergei Odintsov, Marek Olechowski, and Yuri Shtanov for discussions.

\end{document}